\documentstyle[prl,graphicx,aps]{revtex}
\begin{document}
\draft
\twocolumn[\hsize\textwidth\columnwidth\hsize\csname @twocolumnfalse\endcsname
\title{Fresnel filtering in lasing emission from scarred modes of
wave-chaotic optical resonators }
\author{N.~B.~Rex, H.~E.~Tureci, H.~G.~L.~Schwefel, R.~K.~Chang and A.~Douglas~Stone}

\address {Department of Applied Physics, P. O. Box 208284, Yale University,
New Haven, CT 06520-8284}

\date{May 24, 2001}

\maketitle
\begin{abstract} 
We study lasing emission from asymmetric resonant cavity (ARC)
GaN micro-lasers. By comparing far-field intensity patterns with images of the
micro-laser we find that the lasing modes are concentrated on three-bounce unstable
periodic ray orbits, i.e. the modes are scarred. The high-intensity emission directions
of these scarred modes are completely different from those predicted by applying Snell's
law to the ray orbit. This effect is due to the process of ``Fresnel filtering'' which
occurs when a beam of finite angular spread is incident at the critical angle for total
internal reflection.
\end{abstract}
 ]

Understanding the correspondence
between classical phase space structures and wave functions for a general
classical dynamics is the goal of investigations in quantum/wave
chaotic systems \cite{Gutzwiller}.  Generic Hamiltonian systems have 
mixed phase spaces which
consist of tori (which support quasi-periodic orbits), stable periodic orbits
with their associated islands of stability, and unstable periodic orbits which
lie in regions of phase space with chaotic motion.  The simplest possibility,
explored in the early days of the field, was that quantum wavefunctions, when
projected into phase space would cover approximately uniformly the various
classical invariant sets;  hence wavefunctions associated with chaotic regions
would fill such regions uniformly (with fluctuations) as would the
corresponding
ergodic classical trajectories.  We now know \cite{Robnik} that this 
situation is only
realized at extremely high quantum numbers and that states localized
on unstable
classical periodic orbits in the chaotic regions of phase space (``scars'') are
common in many systems of interest \cite{scars1,scars2}.  The same
considerations which
lead to scarred eigenstates of the Schr\"odinger equation also imply
that the wave
equation of electromagnetism will have scarred modes when its boundary
conditions (e.g. shape of a resonator) generate chaotic ray motion; and indeed
such modes have been previously observed in microwave cavities \cite{microwave}.

It has been shown that dielectric {\it optical} micro-cavities and
micro-lasers represent a realization of a wave-chaotic system and one that
presents many unsolved problems for optical
physics \cite{Mekis,nature,science,PRL,austria}.  For example,
quadrupole-deformed InGaAs and GaAs quantum cascade micro-lasers which lased on stable bow-tie
modes were found to produce 1000 times higher output power than undeformed cylindrical
lasers of the same type which lased on whispering gallery
modes \cite{science,austria}.  The mechanism
of mode selection and the increase of output power in these devices is not
currently understood. Recently we reported preliminary data \cite{nathan} 
indicating that in deformed GaN diode lasers the stable bow-tie modes are not
selected but instead unstable ``triangle'' modes are the dominant
ones.  This was
the first time that scars had been observed in an active as opposed to a
passive cavity.
Moreover, while these modes are based on triangular periodic orbits
which strike the boundary near
the critical angle for total internal reflection, their emission
intensity pattern is completely
different from that expected by applying Snell's law to the
underlying periodic orbit.  Below we
interpret this surprising finding as due to an effect we term ``Fresnel
Filtering'' which arises when a beam of finite angular spread is partially
transmitted through a dielectric interface near the total internal reflection
condition.  This is a generic effect, somewhat similar to the well-studied
Goos-H\"anchen shift for a reflected beam \cite{GH1,GH2}, but it has
not to our knowledge been
clearly identified or studied in the optics literature.  We are able to
establish this effect because, in contrast to earlier studies, we
simultaneously collect far-field emission patterns {\it and} images of the
sidewall of the resonator.  Two other groups have very recently reported
lasing emission from
dielectric micro-cavities which they interpret as due to scarred
modes \cite{Narimanov,Lee}.

\begin{figure}[b]
\centering
\includegraphics[height=185mm]{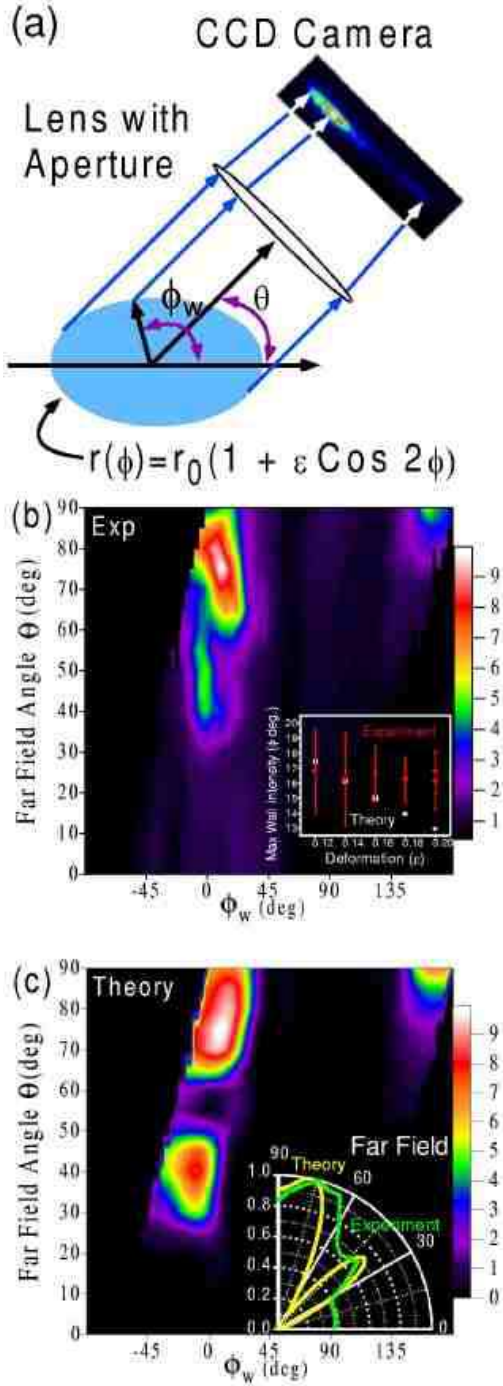}
\caption{(a) Experimental setup for measuring simultaneously
far-field intensity patterns and
images of the sidewall emission. (b) Experimental data showing in
color scale the CCD
images (converted to sidewall angle $\phi_W$) as a function of camera
angle $\theta$.  Three bright
spots are observed on the boundary for camera angles in the 1st
quadrant, at $\phi_W \approx
17^{\circ}, 162^{\circ}, -5^{\circ}$.  Inset shows the position of
the bight spot in the 1st quadrant vs.\ deformation, compared to the location
of the triangular periodic orbit
(see insets Figs.~\ref{fig:husimi}(a), \ref{fig:filter}(a)). (c)
Calculation of expected image data using the scarred mode shown in
Fig.~\ref{fig:husimi}(a); inset shows calculated
and experimental far-field patterns obtained by integrating over
$\phi_W$ for each $\theta$.}
\label{fig:setup}
\end{figure}

\begin{figure}[h]
\centering
\includegraphics[width=6cm]{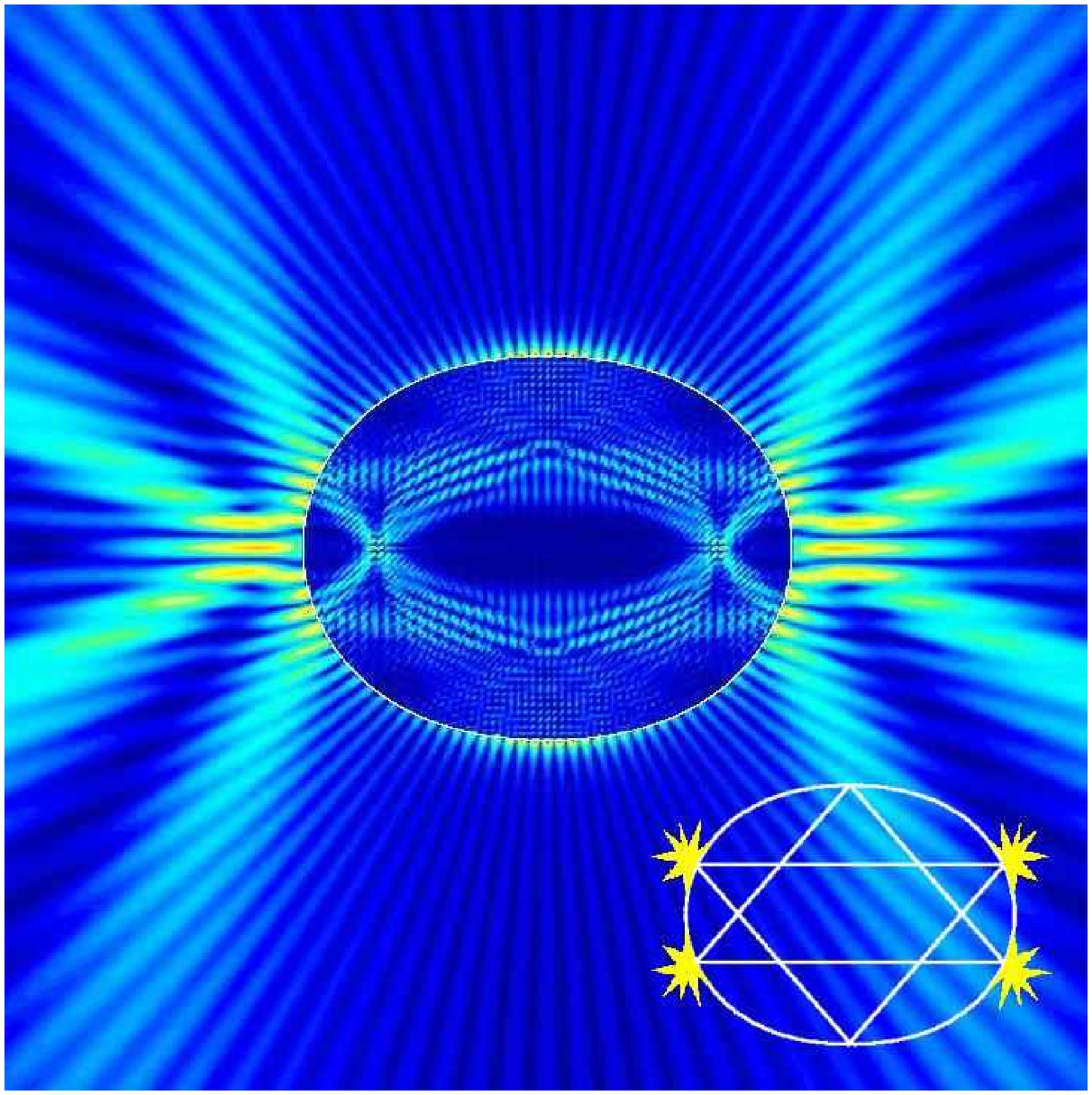}
\centering
\includegraphics[width=\linewidth]{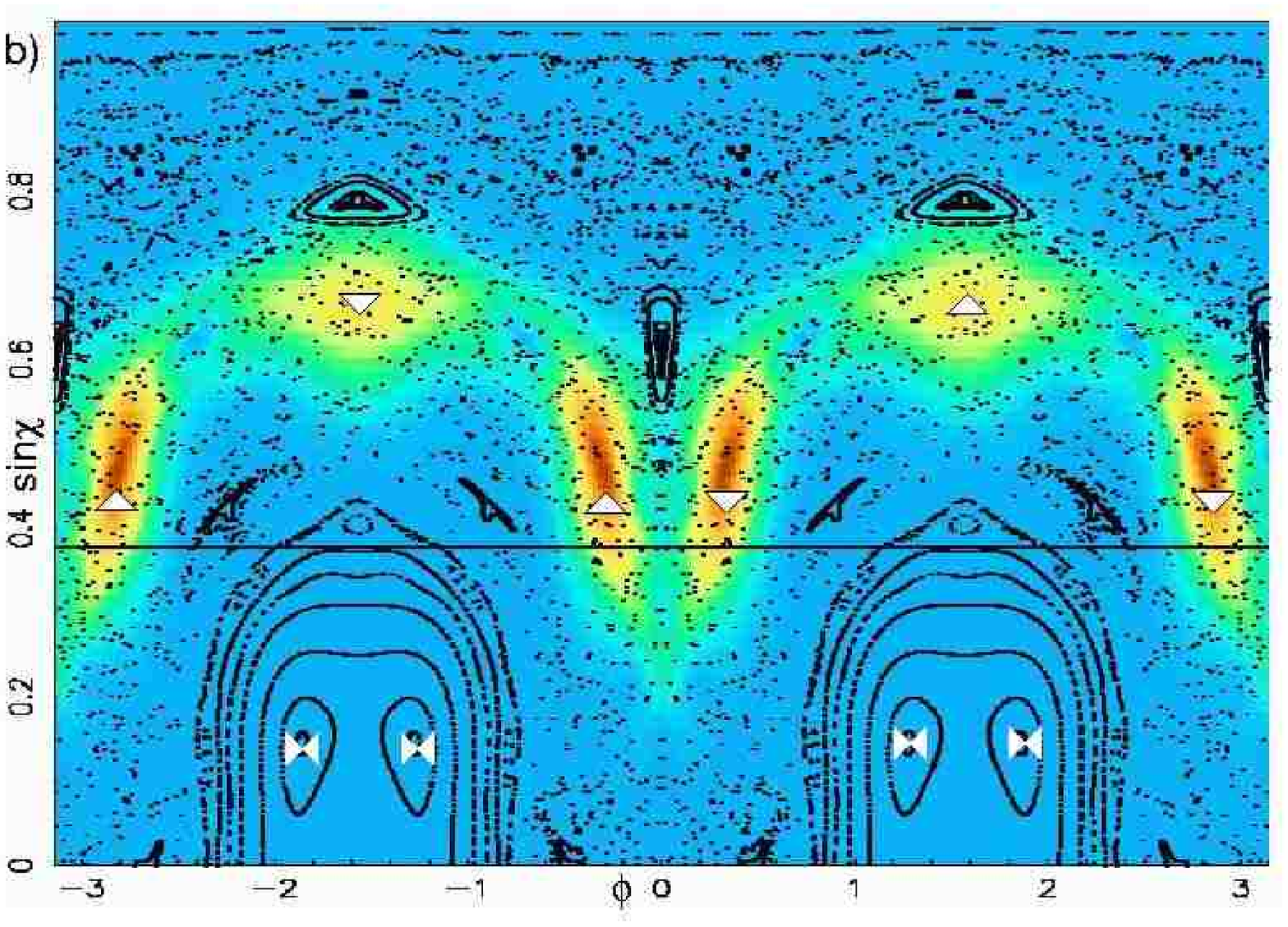}
\caption{(a) Real-space false color plot of the modulus of the
electric field for a calculated
quasi-bound state of $nkr_0=129$ (n is the index of refraction, $k=2\pi/\lambda$)
and $\epsilon = 0.12$ which scars
the triangular periodic orbits shown in the inset (M. V. Berry has termed this the 
``Scar of David'').
The four points of low
incidence angle which should emit strongly are indicated. (b) Husimi
(phase-space distribution) for
the same mode projected onto the surface of section of the resonator.
The x-axis is $\phi_W$ and the
y-axis is
$\sin \chi$, the angle of incidence at the boundary. The surface of
section for the corresponding ray
dynamics is shown in black, indicating that there are no stable
islands (orbits) near the high
intensity points for this mode.  Instead the high intensity points
coincide well with the bounce
points of the unstable triangular orbits (triangles).  The black line
denotes $\sin \chi_c = 1/n$ for
GaN; the triangle orbits are just above this line and would be
strongly confined whereas the stable
bow-tie orbits (bow-tie symbols) are well below and would not be
favored under uniform pumping
conditions.}
\label{fig:husimi}
\end{figure}

The experimental set-up is shown schematically in 
Fig.~\ref{fig:setup}(a).  A GaN
micro-laser of
refractive index $n=2.65$ is optically pumped at 355nm and emits at
375nm.  The
structure is based on MOCVD-grown GaN on a sapphire substrate which is etched
from a mask using standard photolithography to create a $2 \mu
\rm{m}$ high pillar
with a quadrupolar deformation of the cross-section, $r(\phi) = r_0(1
+\epsilon \cos 2
\phi)$ with $r_0 = 100 \mu \rm{m}$.  Light emitted from the laser is imaged
through an aperture subtending a $5^{\circ}$ angle and lens onto a
CCD camera which is rotated by an
angle $\theta$ in the far-field from the major axis.  A bandpass filter
restricts the imaged light to the stimulated emission region of the GaN
spectrum.  The CCD camera records an image of the intensity profile on the
sidewall of the pillar as viewed from the angle $\theta$ which is
converted from pixels to angular position $\phi_W$. Summing these
intensities yields the total far-field intensity emitted in direction $\theta$.
Data were taken for quadrupole lasers with $\epsilon = 0.12, 0.14,
0.16, 0.18$, and $0.20$ and for other shapes as well. The full data set will be
analyzed in a later work, here we
focus on the data for $\epsilon=0.12$
which show a simple scarred mode. In Fig.~\ref{fig:setup}(b) these data are displayed in a color scale
which identifies both the
highest emission directions and the brightest points on the sidewall
(labelled by their angle $\phi_W$).  The data show that the maximum intensity in the 1st
quadrant is observed at angle $\theta \approx 74^{\circ}$ and is emitted from
the region of the sidewall around $\phi_W \approx 17^{\circ} $.  The data also
show a secondary bright spot at slightly negative $\phi_W \approx
-5^{\circ} $ and another one at
$\phi_W \approx 162^{\circ}$.  The observation of a small number of
well-localized
bright spots on the sidewall suggests a lasing mode based on a short periodic
ray trajectory.  The two-bounce stable Fabry-Perot mode would emit
from $\phi_W =
90^{\circ}$ in the direction $\theta=90^{\circ}$.  The stable four-bounce
bow-tie mode, dominant in the devices of Ref.~\cite{science}, is also
inconsistent with our data.  It is very low-Q at this deformation due 
to its small
angle of incidence and would give bright spots at $\phi_W = 
90^{\circ} \pm 17^{\circ}
$, far away from the brightest spot at $\phi = 17^{\circ}$.  There is however a
pair of symmetry-related isosceles triangular orbits (inset,
Fig.~\ref{fig:husimi}(a)) with
bounce points very close to the observed bright spots (see inset to
Fig.~\ref{fig:setup}(b)).
These orbits are unstable for $\epsilon > 0.098 $ with trace of the monodromy
matrix equal to -5.27 at $\epsilon=0.12$. The two equivalent bounce points in
each triangle at $\phi_W= \pm 17^{\circ}$ and $180^{\circ} \pm 17^{\circ}$ have
$\sin \chi \approx 0.42$, very near to the critical value, $\sin \chi_c=1/n=
0.38$, whereas the bounce points at $\phi_W = \pm 90^{\circ}$ have $\sin \chi =
0.64$ and should emit negligibly (inset to Fig. \ref{fig:husimi}(a)).
This accounts for the three
bright spots observed experimentally in Fig.~\ref{fig:setup}(b) (the
fourth spot at $\phi_W \approx
197^{\circ}$ does not emit into the first quadrant, see also inset to
Fig. 3(a)).  Solutions for the
quasi-bound states of this resonator in the empty cavity can be found
numerically, both in
real-space and phase space, and we find that indeed there exist such scars (see
Figs.~\ref{fig:husimi}(a), (b)).  Here we plot both the modulus of the
electric field in real-space
and the projection of the Husimi distribution of the mode onto the surface
of section of the resonator \cite{Frischat}.  The Husimi distribution is
a (gaussian) smoothed version of the Wigner transform of the mode, which
represents a wavefunction or mode as a phase space density consistent with the
uncertainty principle.  Projection onto the surface of section then gives a
measure of the density of rays which strike the boundary at a given position,
$\phi_W$ and a given incidence angle, $\chi$.
Additionally we evaluate this mode in the far-field
and find an emission pattern in good agreement with the experimental
measurement (see inset, Fig.~\ref{fig:setup}(c)).
Finally, if we take this scarred mode and propagate it numerically 
via a lens transform
\cite{Goodman} we obtain the result
shown in Fig.~\ref{fig:setup}(c), which is in quite reasonable
agreement with the experimental data
of Fig.~\ref{fig:setup}(b), taking into account that the lasing
mode should differ somewhat from
the resonance of the empty cavity.  Hence we conclude that the
dominant lasing mode in the experiment is such a scarred mode.

\begin{figure}[ht]
\centering
\includegraphics[width=\linewidth]{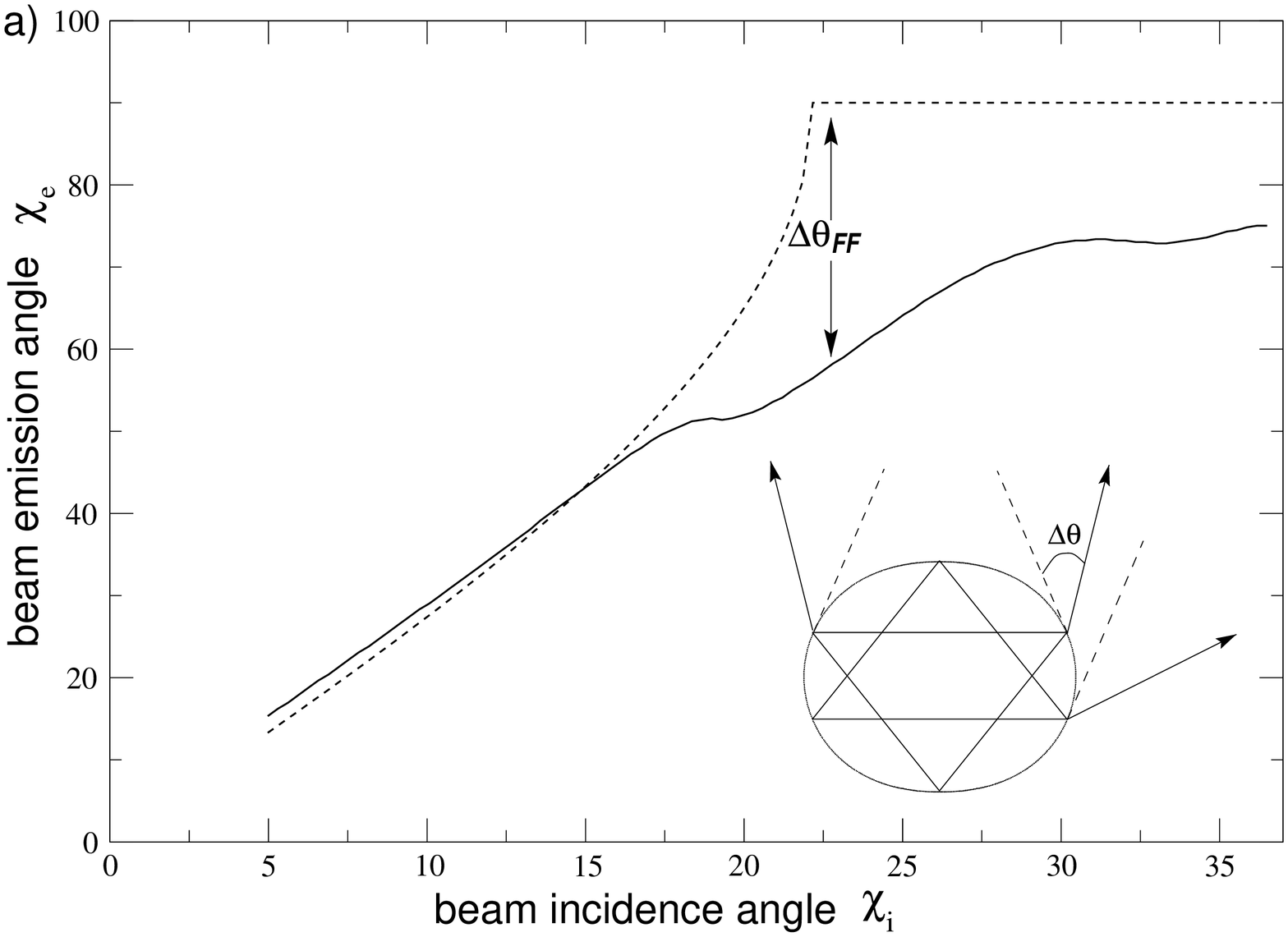}
\centering
\includegraphics[width=\linewidth]{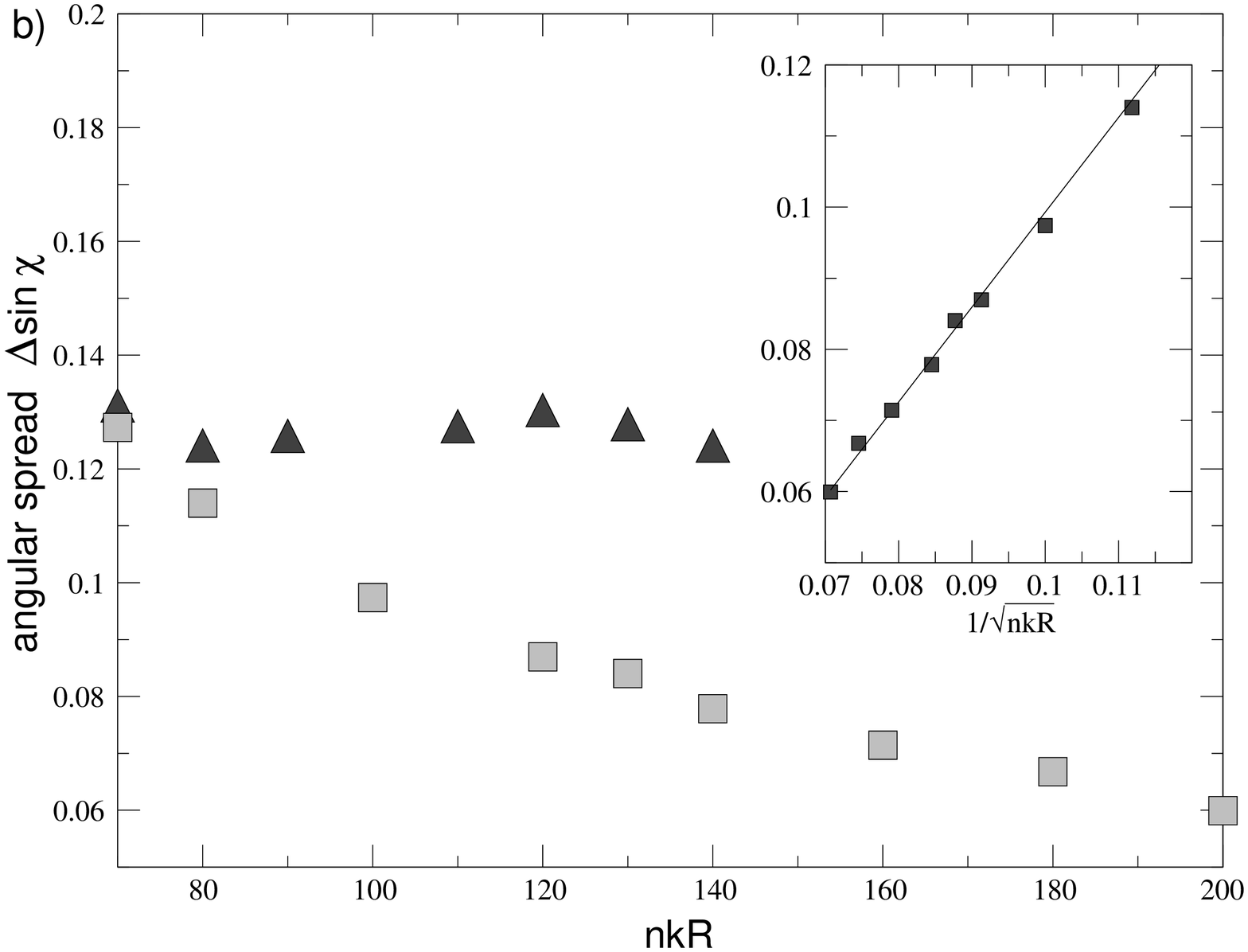}
\caption{(a) Solid line: beam emission angle vs.\ incidence angle for a
beam of angular
spread equivalent to the scarred mode of Fig.~\ref{fig:husimi} 
incident on a plane
interface.  Dashed line
is Snell's law, the discrepancy is the Fresnel Filtering angle
$\Delta \theta_{FF}$.  Inset
schematic shows the three emitted ``beams'' detected in the experiment
and illustrates their
strong deviation from Snell's law (dashed tangent lines). (b)
Dependence of angular spread
of the ``incident beams'' vs. $nkr_0$ for scarred triangle modes
(triangles) and stable
(gaussian) bow-tie modes; inset shows that the spread decreases as
$1/\sqrt{nkr_0}$ (see inset) for stable gaussian
modes as predicted, whereas no clear variation with $nkr_0$ is seen
for the scarred modes.}
\label{fig:filter}
\end{figure}

The data of Figs.~\ref{fig:setup}(a), (b) however present an intriguing
puzzle from the point of
view of ray optics.  A mode localized on these triangular orbits would be
expected to emit from the four bounce points approximately in the tangent
direction according to Snell's law; this means that the bright spot
at $\phi_W =
17^{\circ}$ should emit into the direction $\theta \approx
115^{\circ}$ (Fig.~\ref{fig:filter}(a)),
whereas the data clearly indicate that the $17^{\circ}$ bright spot
emits in the
direction $\theta = 72^{\circ}$.  Thus the emission pattern violates the
intuitive expectations of ray optics by $43^{\circ}$, a huge discrepancy (see
Fig.~\ref{fig:filter}).  Moreover the ratio $\lambda /nR =
2.8\times10^{-3}$, so we are far into
the regime in which the wavelength is small compared to the geometric features
of the resonator and ray optics would be expected to be a good
approximation.  The resolution of
this apparent paradox is suggested by the numerical data of
Fig.~\ref{fig:husimi}(b).  It is clear
that the scarred mode, while localized around the triangle orbit, has
a significant
spread in angle of incidence, $\Delta \sin \chi \approx 0.2$.  This means that
we must regard the scarred mode as a (non-gaussian) beam with a large
angular spread, with some
components almost totally reflected and other components transmitted
according to the Fresnel
transmission law.  This law is rapidly varying near $\sin \chi_c =
1/n$ and strongly
favors the components with lowest angle of incidence.  Thus the
centroid of the transmitted
beam is shifted away from the tangent direction (i.e. away from the geometric
optics result for a plane wave) by an amount $\Delta \theta_{FF}$ by
this effect,
which we call ``Fresnel filtering''.  Since the sidewall of the resonator is
curved, strictly speaking the Fresnel law (which is for an infinite flat
surface) does not hold, but the curvature corrections should be small
when $\lambda/nR
\ll 1$, as is the case here.  Thus we can evaluate the Fresnel filtering effect
theoretically for a flat dielectric interface.
We express the outgoing beam in the
angular spectrum representation\cite{Wolf} and evaluating the 
resulting integral
by the saddle point method in the asymptotic farfield region, we obtain 
the far-field intensity pattern $I(\theta)$ \cite{Hakan}:
\begin{eqnarray*}
\lefteqn{I(\theta)  \propto {\cal T}\left(\frac{\sin 
\chi_e(\theta)}{n}\right) {\cal
P}\left(\frac{\sin\chi_e(\theta)}{n}\right)} \\
&& \times \frac{\sin\chi_e(\theta)
\sin\chi_{i}+\cos\chi_{i}\sqrt{n^{2}-\sin^{2}\chi_e(\theta)}}
{\sqrt{n^{2}-\sin^{2}\chi_e(\theta)}} \cos\chi_e(\theta)
\end{eqnarray*}
where ${\cal P}(x)$ is the probability density of sine of the incidence angle
for the beam
and ${\cal T}(x) =  \sqrt{1 - x^2}/(\sqrt{1 - x^2} + \sqrt{1 - n^2x^2})$.
For the plane interface $\chi_e(\theta) = \theta$, while for the equivalent 
resonator $\chi_e = \theta - \cos^{-1} [ \hat{n}
(\phi_b) \cdot \hat{x}]$, where $\hat{n} (\phi_b)$ is the unit
normal at the bounce point.

To model the experiment we assume that the probability distribution for
the incidence angle is approximately the same as the cross-section of
the Husimi distribution of Fig.
\ref{fig:husimi}(b) evaluated at the triangle bounce point $\phi_W = 
17^{\circ}$.
In Fig.~\ref{fig:filter} we plot the beam emission angle $\chi_e(\theta)$ defined as
the angular maximum of the far-field pattern vs.\ incidence angle $\chi_i$. We find a
very large angular shift
$\Delta \theta_{FF}$, in approximate agreement with experiment (we don't expect
precise quantitative agreement since the curvature of the resonator also
has some effect on the far-field pattern, giving a different value of $\theta_{FF}$
than for the planar interface).

The size of the Fresnel filtering effect depends strongly on the
angular beam spread.
For gaussian resonator modes one can show that this spread
tends to zero as $1/\sqrt{nkr_0}$ \cite{Hakan}, (see inset 
Fig.~\ref{fig:filter}(b)).
Since our numerical simulations of the scarred
mode are for $nkr_0 = 129$, whereas
the experiment corresponds to $nkr_0 \approx 4,440$, one may ask
whether the large Fresnel Filtering
angle found in Fig.~\ref{fig:filter}(a) (for $nkr_0 = 129$) will 
extrapolate correctly
to agree with the
experiment.  As there is currently no theory of this scaling for scarred modes,
we studied the
scaling of the angular width numerically (Fig.~\ref{fig:filter}(b)). 
We found no detectable
decrease in the angular width with $nkr_0$, in clear contrast to the
behavior of the gaussian
modes. Thus, while the Fresnel Filtering effect should be present for
gaussian resonator modes based on
stable periodic orbits, it appears to be significantly enhanced for
unstable (scarred) modes.

In conclusion, we have found that the dominant lasing mode in
quadrupolar GaN micro-lasers
are unstable (scarred) modes. For resonators with chaotic ray
dynamics, such scarred modes play
a special role as they allow high-Q resonances despite the ray chaos.
Such modes exhibit a novel
emission pattern, which is completely different from that expected by
applying Snell's law
to the underlying periodic ray trajectory, due to the phenomenon of
Fresnel Filtering. 

We acknowledge helpful discussions with P. Jacquod. This work was supported by NSF 
grants DMR-0084501, PHY-9612200 and AFOSR grant F49620-00-1-0182-02.

\end{document}